\documentclass[usenatbib]{mn2e}

\newif\ifAMStwofonts

\title[Multiresolution approach for period determination on unevenly sampled data]{Multiresolution approach for period determination on unevenly sampled data}
\author[X. Otazu et al.]
{X. Otazu$^{1}$\thanks{E-mail: xotazu@cvc.uab.es (XO);
mribo@discovery.saclay.cea.fr (MR);
jmparedes@ub.edu (JMP);
marta.peracaula@udg.es (MP);
jorge@am.ub.es (JN)},
M. Rib\'o$^{2\star}$, 
J. M. Paredes$^{3\star}$\thanks{CER on Astrophysics, Particle Physics
and Cosmology. Universitat de Barcelona},
M. Peracaula$^{4\star}$
and J. N\'u\~nez$^{3,5\star}$\\
$^1$Centre de Visi\'o per Computador, Campus Universitat Aut\`onoma de Barcelona, 08193 Cerdanyola del Vall\`es (Barcelona), Spain \\
$^2$Service d'Astrophysique, CEA Saclay, B\^at. 709, L'Orme des Merisiers, 91191 Gif-sur-Yvette, Cedex, France \\
$^3$Departament d'Astronomia i Meteorologia, Universitat de Barcelona, Av. Diagonal 647, 08028, Barcelona, Spain \\
$^4$Institut d'Inform\`atica i Aplicacions, Universitat de Girona, Campus de
Montilivi s/n, 17071 Girona, Spain\\
$^5$Observatori Fabra, Cam\'{\i} de l'Observatori sn, 08035 Barcelona, Spain}

\date{Accepted YYYY MMMMMMMM DD.
      Received YYYY MMMMMMMM DD}

\pagerange{\pageref{firstpage}--\pageref{lastpage}}
\pubyear{YYYY}

\begin{document}

\maketitle

\label{firstpage}

\begin{abstract}
In this paper we present a multiresolution-based method for period
determination that is able to deal with unevenly sampled data. This method
allows us to detect superimposed periodic signals with lower signal-to-noise
ratios than in classical methods. This multiresolution-based method is a
generalization of the wavelet-based method for period detection that is unable
to deal with unevenly sampled data, presented by the authors in \cite{otazu}.
This new method is a useful tool for the analysis of real data and, in
particular, for the analysis of astronomical data.
\end{abstract}


\section{Introduction}
\label{intro}

The detection of periodic signals in astronomical data has been usually
addressed by classical Fourier-based or epoch folding methods. These methods
have different problems when dealing with non-sinusoidal periodic signals or
with very low signal-to-noise ratios. When the analysed data set contains
several periodic signals, the behavior of classical period determination
methods highly depends on intrinsic signal characteristics (see, for example,
\citealt{lafler}; \citealt{jurkevich}; \citealt{stellingwerf}; \citealt{lomb};
\citealt{scargleA}, \citeyear{scargleB}; \citealt*{roberts}; \citealt{press}).
The reader is referred to the introduction of \cite{otazu} (hereafter Paper~I)
for a general discussion. To avoid this problem, we presented there, as a
preliminary step, a wavelet-based approach that only works with evenly time
sampled data. 

In astronomy, however, this is not a usual situation, since data are mostly
acquired on irregular intervals of time. In such a case there are two
possibilities: resample the data into a new evenly sampled data set, or use a
method able to deal with the original unevenly sampled data set. In the first
case we are forced to modify the original data, which necessarily implies a
loss of information. Moreover, this is not always possible if the temporal
gaps are larger than some of the periods present in the data. In order to
avoid these problems, a technique capable to deal with unevenly sampled data
is needed.


The present paper is a natural extension of Paper~I that allows to work on
unevenly time sampled data. We show how the methodology of multiresolution
decomposition (similar to the wavelet decomposition philosophy) is very well
suited to this problem, since it is completely oriented towards decomposing functions into several frequential characteristics.

As in Paper~I, the main objective is to isolate every signal present in our
data and to analyse them separately, avoiding their mutual influences. In
Section~\ref{multiresolution} we outline some concepts in multiresolution
analysis and their similarities with wavelet theory that are relevant to the
stated problem. In Section~\ref{period} we propose an algorithm to detect each
of the periodic signals present in a data set by combining multiresolution
analysis decomposition with classical period determination methods. In
Sections~\ref{simulated} and \ref{results} we present some examples of
synthetic data we used to test the algorithm and the results we obtained. We
summarise our conclusions in Section~\ref{conclusions}.

\section{Multiresolution analysis}
\label{multiresolution}

Multiresolution decomposition introduces the concept of the presence of
details between successive levels of resolution. Many wavelet decomposition
algorithms are based on multiresolution analysis schemes (\citealt{chui};
\citealt{daubechies}; \citealt{meyer}; \citealt{kaiser}; \citealt{vetterli};
\citealt{mallat}; \citealt{starck}), and some astronomical applications using
wavelets for timing analysis have been reported (\citealt*{szatmary};
\citealt{foster}; \citealt{ribo}; \citealt{poly}). In fact, all these methods
use the same philosophy and the obtained results can be interpreted in the
same way. However, wavelet theory presents some constraints on mathematical
functions. These constraints are not respected in all multiresolution
decomposition schemes. Therefore, when we are using a certain multiresolution
decomposition algorithm that fulfills the wavelet constraints, we are
obtaining a wavelet decomposition. In contrast, when a given multiresolution
decomposition algorithm violates these wavelet constraints, we are obtaining a
result similar to, but which is not, a wavelet decomposition.

As discussed in Paper~I, we note that there are wavelet approaches that are
based on approximations of a continuous wavelet transform and on the
subsequent study of wavelet space coefficients (see, e.g.,
\citealt{szatmary}), which are able to deal with unevenly sampled data sets.
However, these algorithms present a {\it non-direct} inverse wavelet
transform, in the sense that the search for periodicities is based on the fit
between the wavelet base function profile and the signal one, and therefore on
the values of the wavelet transform coefficients, which highly depend on the
wavelet base used. Moreover, the period analysis has to be performed on the
wavelet coefficients space but, since it is usually decimated, accurate period
detection is a difficult task.
The multiresolution decomposition scheme we use in this work performs the
decomposition on the temporal space, which allows to find accurate values for
periodicities.


\subsection{A multiresolution analysis algorithm}
\label{multires_alg}

In order to obtain a multiresolution decomposition for signals, an algorithm
to decompose the signal into frequency planes can be defined as follows. Given
a signal $p$ we construct the sequence of approximations:
\begin{equation}
p_1 = F_1(p),~~~~~
p_2 = F_2(p_1),~~~~~
p_3 = F_3(p_2),\cdots,
\end{equation}
performing successive convolutions with Gaussian filters $F_i$.

It is important to note the difference between this sequence of convolutions
and the one used in Paper~I. In the latter we were dealing with a discrete
convolution mask, and hence forced to work with evenly spaced data. In
contrast, the continuous nature of the convolution functions used in the
present paper, allows us to work with unevenly sampled data.

Similarly to the wavelet planes, the multiresolution frequency planes are
computed as the differences between two consecutive approximations $p_{i-1}$
and $p_i$. Letting $w_i = p_{i-1} - p_i \;\;(i=1,\cdots,n)$, in which $p_0 =
p$, we can write the reconstruction formula: \begin{equation} p =
\sum_{i=1}^{n}w_i + p_r\;\;. \label{wav_des} \end{equation}

In this representation, the signals $p_i~(i=0,\cdots,n)$ are versions of the
original signal $p$ at increasing scales (decreasing resolution levels),
$w_i~(i=1,\cdots,n)$ are the multiresolution frequency planes and $p_r$ is a
residual signal (in fact $n=r$, but we explicitly substitute $n$ by $r$ to
clearly express the concept of $residual$). In our case, we are using a dyadic
decomposition scheme. This means that the standard deviation of the Gaussian 
function associated with the $F_i$ filter is $\sigma_i=2\sigma_{i-1}$. Thus,
similarly to the wavelet approach, the original signal $p_0$ has double
resolution compared to $p_1$, and so on. All these $p_i~(i=0,\cdots,n)$ signals
have the same number of data points as the original signal $p$.

Since $\sigma_{i+1}$ depends on $\sigma_i$, a value for $\sigma_0$ has to be
carefully chosen for every data set. It has to be fixed considering a likely
minimum value for the time-duration of the features and the characteristic time
sampling of the data set, in order to include a significant number of points on
which to perform the convolutions. Too small values do not accurately describe
feature profiles, and suffer from poor or noisy data. In contrast, too large
values reduce the noise effect by integrating a lot of data points, but may
ignore interesting high frequency features. Hence, a first analysis of data has
to be performed in order to obtain a useful initial $\sigma_0$ value.

We have used the same notation as in the wavelet decomposition described in
Paper~I because, as explained above, the idea of these multiresolution planes
is similar to the wavelet ones.

We note that this particular decomposition scheme that uses a Gaussian kernel
can also be interpreted as a scale-space filtering (\citealt{baubaud},
\citealt{sporring}, \citealt{witkin94}) or as a particular case of more
general image diffusion approaches (\citealt{lindeberg}, \citealt{perona}).
Smoothed data sets $p_i$ can be interpreted as diffused scale-space images,
and the difference between them as the details at different scales.

\section{Period detection algorithm}
\label{period}

We propose to apply this multiresolution analysis algorithm to solve our
initially stated problem: to isolate each of the periodic signals contained in
a set of unevenly sampled data and study them separately.

In order to do so, we proceed as in Paper~I, and the period detection
algorithm we propose is as follows:

\begin{enumerate}

\item choose values for $\sigma_0$ and $n$; decompose the original signal $p$
into its multiresolution frequency planes $w_i~(i=0,\cdots,n)$.

\item detect periods in each of the $n$ obtained frequency planes $w_i$.

\end{enumerate}

Phase Dispersion Minimization ({\sevensize PDM}) \citep{stellingwerf} and
{\sevensize CLEAN} \citep{roberts} methods are used to detect periods in the
original data and in every multiresolution frequency plane. In Paper~I we
described several undesirable effects of {\sevensize PDM} and {\sevensize
CLEAN} methods on data with superimposed signals, as well as the advantages of
using multiresolution-based methods over the classical ones.

Hereafter, and for notational convenience, the multiresolution-based
{\sevensize PDM} and {\sevensize CLEAN} methods will be called {\sevensize
\mbox{MRPDM}} and {\sevensize MRCLEAN}, respectively.

\section{Simulated data}
\label{simulated}

In order to check the benefit of applying {\sevensize MRPDM} versus
{\sevensize PDM}, or {\sevensize MRCLEAN} versus {\sevensize CLEAN}, we
proceeded similarly as in Paper~I by generating and analysing several sets of
simulated data containing two superimposed periodic signals. Each data set is
composed of a high-amplitude primary sinusoidal function and a secondary
low-amplitude Gaussian one. Finally, we added a white-Gaussian noise to this
combination of signals. We increased the value of the noise standard
deviation, $\sigma$, up to the value where detection of periodic signals
became statistically insignificant in both the classical and the
multiresolution-based methods.

The primary signal is intended to simulate variable sources with a pure
sinusoidal intensity profile (like precession of accretion discs), and the
secondary, burst-like events (like pulses or eclipses) superimposed to it. We
note that in Paper~I we also studied the superposition of two sinusoidal
functions, the so called sine+sine case. However, here we have directly
focused on the more realistic situation of the so called sine+Gaussian case,
where the difference between the classical and the multiresolution-based
methods is more critical (see Paper~I).

The characteristics of the signals are:

\begin{enumerate}

\item Each signal is generated as an unevenly spaced data set. Its time
sampling has been taken from observations of the X-ray binary LMC~X-1 by the
All Sky Monitor onboard the RXTE satellite (\citealt{levine}). We have used an
observation period which has 8270 measurements during 679 days.

\item The high-amplitude sinusoidal function has an amplitude equal to 1 and a
period of 108.5 days.

\item The amplitudes of the low-amplitude periodic Gaussian function are 0.1
and 0.5.

\item The periods used for the secondary function are 13.13 and 23.11 days.

\item We have used two values for the full width at half-maximum (FWHM) of the
Gaussian signal, corresponding to 2 and 6 days, respectively.

\end{enumerate}

In the first four columns of Table~\ref{gauss_table} we present the parameters
used to generate each simulated data set.

\section{Results}
\label{results}

We recall that the simultaneous use of two independent methods, such as
{\sevensize CLEAN} and {\sevensize PDM}, is usually applied to discriminate
false period detections from the true ones. A similar procedure can be used
with each one of the multiresolution planes in the {\sevensize MRPDM} and
{\sevensize MRCLEAN} methods. Therefore, when comparing the behavior of these
different methods, we have to compare the usual {\sevensize PDM-CLEAN} method
combination for period estimation prior to the new {\sevensize MRPDM-MRCLEAN}
combination.

Taking into account the characteristics of the simulated data we have chosen a
$\sigma_0=1$~day. As in Paper~I, the primary period (108.5 days) is always
detected by all methods, and does not appear in Table~\ref{gauss_table}. In
this table, the four last columns show the detected low-amplitude periods for
each data set using {\sevensize PDM}, {\sevensize CLEAN}, {\sevensize MRPDM}
and {\sevensize MRCLEAN} methods, respectively. A dash is shown when a period
is not detected, and a question mark when the detection is difficult or
doubtful. When a period is found in the multiresolution-based methods, we also
show in parentheses the mutiresolution planes where it is detected.

\begin{table*}
\centering 
 \begin{minipage}{133mm}
\caption[]{\label{gauss_table} Periods, in days, detected in the data sets. The first four columns are the parameters used for the simulated data. A dash is shown when a period is not detected, and a question mark when the detection is difficult or doubtful. When a period is indicated in the multiresolution-based methods, we also show in parentheses the multiresolution planes where it is detected (using $\sigma_0=1$~day).}
\begin{tabular}{ccclllll}
\hline
FWHM & Period & Amplitude & $\sigma_{\mathrm noise}$ & {\sevensize PDM} &
{\sevensize CLEAN} & {\sevensize MRPDM} & {\sevensize MRCLEAN} \\
\hline

2 & 	13.13		        &	0.1	&	       	0.0	&	-	&	13.13		&13.13 (1,2)	&	13.13 (1,2) \\
 & 		&&						0.1	&	-	&	13.13		&13.13 (1,2)	&	13.13 (1,2)\\
 & 		&&						0.15	&	-	&	13.13		&13.13 (1?,2)	&	13.13 (1?,2)\\
 & 		&&						0.2	&	-	&	13.15?	&13.13 (1?,2)	&	13.13 (2)\\
 & 		&&						0.25	&	-	&		-		&13.14 (1?,2)	&		-\\
 & 		&&						0.3	&	-	&		-		&13.14 (2)		&		-\\
 & 		&&						0.35	&	-	&		-		&13.14 (2?)		&		-\\
 & 		&&						0.4	&	-	&		-		&	-				&		-\\
 &  		&& & & & & \\
 & 		&			0.5	&		0.0	&	-	&	13.13		&13.13 (1,2,3)		&	13.13 (1,2,3)\\
 & 		& &						0.5	&	-	&	13.13		&13.13 (1?,2,3)	&	13.13 (1?,2,3)\\
 & 		& &						1.0	&	-	&	13.14?	&13.13 (2,3)		&	13.13 (2,3)\\
 & 		& &						2.0	&	-	&		-		&13.14 (2,3)		&	13.14 (3?)\\
 & 		& &						2.25	&	-	&		-		&13.14 (2?)			&		-\\
 & 		& &						2.5	&	-	&		-		&	-					&		-\\
 & 		&& & & & & \\

 & 	23.11		        &	0.1	&		0.0	&	-	&	23.13		&23.11 (1,2?)	&		-\\
 & 		& &						0.1	&	-	&	23.19		&23.11 (2?)		&		-\\
 & 		& &						0.15	&	-	&		-		&	-				&		-\\
 & 		&& & & & & \\

 & 		&			0.5	&		0.0	&	-	&	23.12		&23.11 (2?,3,4)	&23.11 (3?)\\
 & 		& &						0.5	&	-	&	23.11?	&23.11 (2?,3,4)	& -\\
 & 		& &						1.0	&	-	&		-		&23.11 (2?,3,4)	& -\\
 & 		& &						1.25	&	-	&		-		&23.11 (2?,3,4)	& -\\
 & 		& &						1.5	&	-	&		-		&	-					& -\\
 & 		&& & & & & \\

6 & 	13.13	        	&	0.1	&		0.0	&	-	&	13.13		&13.13 (1,2,3)		&	13.13 (1,2,3)\\
 & 	& &							0.1	&	-	&	13.14		&13.13 (1,2,3)		&	13.13 (1,2,3)\\
 & 	& &							0.2	&	-	&	13.14		&13.13 (1?,2,3)	&	13.13 (1?,2,3)\\
 & 	& &							0.3	&	-	&	13.14		&13.13 (2,3)		&	13.13 (2,3)\\
 & 	& &							0.4	&	-	&	13.15?	&13.14 (2,3)		&	13.14 (3)\\
 & 	& &							0.6	&	-	&		-		&13.14 (2,3)		&	13.14 (3?)\\
 & 	& &							0.8	&	-	&		-		&13.15 (3)			&	13.15 (3?)\\
 & 	& &							0.9	&	-	&		-		&	-					&	- \\
 & 	&& & & & & \\

 & 	&				0.5	&		0.0	&13.13	&13.13	&13.13 (1,2,3)		&	13.13 (1,2,3)\\
 & 	& &							0.5	&13.13	&13.12	&13.13 (1,2,3)		&	13.13 (1,2,3)\\
 & 	& &							1.0	&13.13	&13.14   &13.13 (1?,2,3)	&	13.13 (1?,2,3)\\
 & 	& &							2.0	&13.13	&13.13	&13.13 (2,3)		&	13.13 (2,3)\\
 & 	& &							3.0	&13.13?	&13.11?	&13.13 (2?,3)		&	13.13 (3)\\
 & 	& &							4.0	&	-		&	-		&	-					&		-\\
 & 	&& & & & & \\

 & 	23.11		        &	0.1	&		0.0	&	-		&23.12		&23.11 (2,3)	   &23.11 (2,3) \\
 & 	& &							0.1	&	-		&23.12		&23.11 (2,3)	   &23.11 (2,3) \\
 & 	& &							0.2	&	-		&23.12		&23.11 (2?,3)      &23.11 (2?,3)\\
 & 	& &							0.3	&	-		&23.12		&23.11 (3)         &23.11 (3)   \\
 & 	& &							0.4	&	-		&23.12?		&23.11 (3?)        &23.11 (3)  \\
 & 	& &							0.5	&	-		&	-	&23.11 (3?)        & - \\
 & 	& &							0.6	&	-		&	-	& -                & - \\
 & 	&& & & & & \\

 & 	& 				0.5	&		0.0	&23.11	&23.11		&23.11 (2?,3,4,5)	& 23.11 (3?,4,5)\\	
 & 	& &							0.5	&23.11	&23.11		&23.11 (2?,3,4,5)	& 23.11 (4,5)\\
 & 	& &							1.0	&23.11	&23.11		&23.11 (3,4,5)		& 23.11 (4,5?)\\
 & 	& &							2.0	&23.11?	&23.12		&23.11 (3?,4,5)	        & 23.11 (4,5?)\\
 & 	& &							2.5	&23.11?	&23.12		&23.11 (4,5?)		& 23.11 (4,5?)\\
 & 	& &							3.0	& -	& -		&23.12 (4)		& 23.11 (4)\\
 & 	& &							3.5	& -	& -		&23.12 (4?)		& - \\
 & 	& &							4.0	& -	& -		& -			& - \\

\hline
\end{tabular}
\end{minipage}
\end{table*}

We must note that the use of two different FWHM for the Gaussian, combined
with two different periods (13.13 and 23.11 days), gives 4 different profiles.
Hence, the phase duration of the burst-like event ranges from very low to
relatively high values in the following order: FWHM=2 and period=23.11, FWHM=2
and period=13.13, FWHM=6 and period=23.11, and finally FWHM=6 and
period=13.13.

In view of the results displayed in Table~\ref{gauss_table} we can make the
following comments:

\begin{enumerate}

\item In all methods, with high noise-to-signal ratios the detected periods
are slightly different from the simulated ones.

\item There is a better performance of {\sevensize CLEAN} over {\sevensize
PDM}. We must note that when the FWHM is only 2 days, {\sevensize PDM} never
detects the secondary period. Only with FWHM=6 days and a relatively high
amplitude (0.5), can {\sevensize PDM} detect the low-amplitude periodic
signals.

\item As the noise increases, the detection starts to fail in the lower
multiresolution planes (higher frequencies), and only the higher ones (lower
frequencies) are noise-free enough to allow period detection.

\item \label{iv} {\sevensize MRPDM} and {\sevensize MRCLEAN} perform better or
similar than {\sevensize PDM} and {\sevensize CLEAN} methods (see exception
below). When {\sevensize CLEAN} marginally detects the secondary period,
{\sevensize MRPDM} and most of times {\sevensize MRCLEAN} have no problems to
detect it, and they work properly even with higher noise. In the {\sevensize
MRPDM} case, the results are always better than with {\sevensize PDM}.

\item In all cases with FWHM=2, the {\sevensize MRPDM} performance is much
better than {\sevensize MRCLEAN}, because the signal is clearly
non-sinusoidal. In the FWHM=6 cases, {\sevensize MRPDM} is only slightly
better than {\sevensize MRCLEAN}, since the signals are closer to a sinusoidal
profile.

\item For a given amplitude of the Gaussian signal, the maximum
noise-to-signal ratio achieved with {\sevensize MRPDM} and {\sevensize
MRCLEAN} increases with the phase duration of the FWHM.

\end{enumerate}

The only exception to \ref{iv} is for the most extreme of the simulated cases,
i.e., the one with the lower amplitude, lower FWHM value and longer period.
However, we note that the period detected by CLEAN is slightly different than
the simulated one.

All these results are very similar to those shown in Table~2 of Paper~I.
Nevertheless the maximum noise-to-signal ratios achieved in the present cases
are around 2.5 times higher. This can be explained because, although the time
span of the data sets used here is around 1.5 times smaller, the number of
points per unit time is around 12 times higher than in Paper~I.

\begin{figure}
\mbox{}
\vspace{6.7cm}
\includegraphics{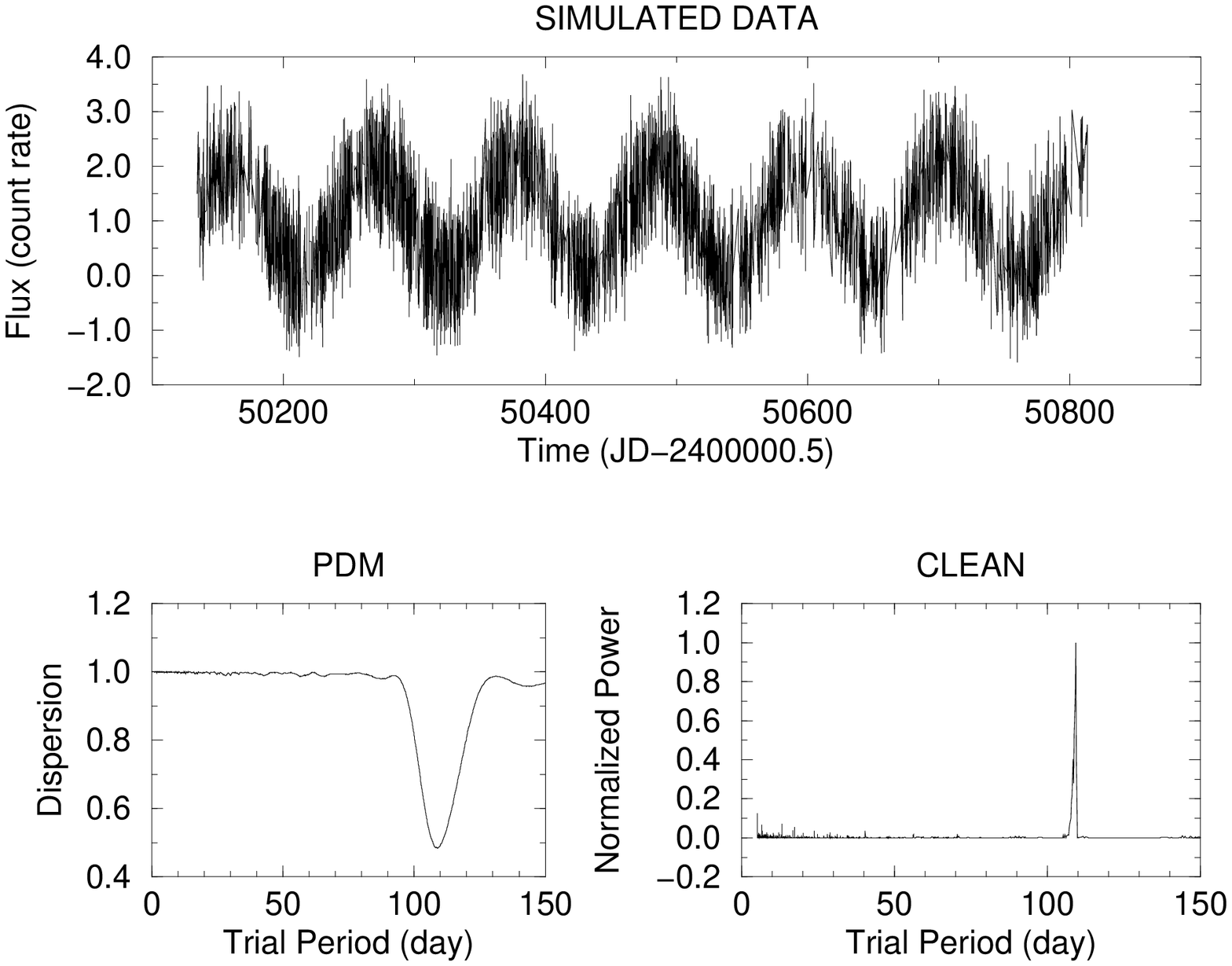}
\caption[]{Top: simulated data of a Gaussian with 13.13 days period, a FWHM of
6.0 days, amplitude=0.1 and $\sigma_{\mathrm noise}=0.6$, superimposed on the sinusoid with 108.5 days period and amplitude=1.0. Bottom: {\sevensize PDM} and {\sevensize CLEAN} outputs of this data set.}
\label{fig:simulated}
\end{figure}

\begin{figure}
\mbox{}
\vspace{5.5cm}
\includegraphics{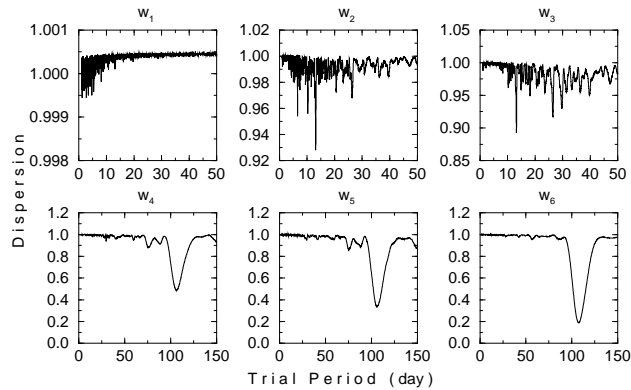}
\caption[]{{\sevensize MRPDM} output for each wavelet plane of the data at the top of Fig.~\ref{fig:simulated}.}
\label{fig:mrpdm}
\end{figure}

\begin{figure}
\mbox{}
\vspace{5.5cm}
\includegraphics{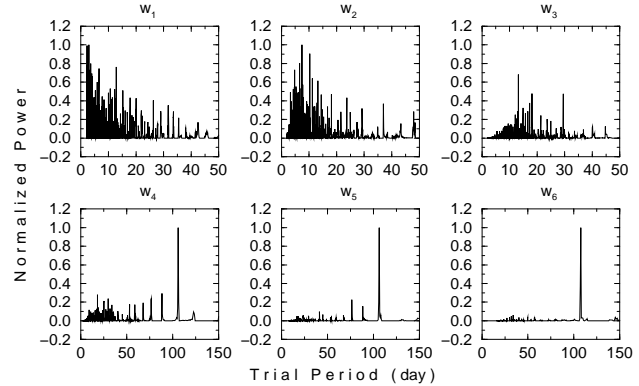}
\caption[]{{\sevensize MRCLEAN} output for each wavelet plane of the data at the top of Fig.~\ref{fig:simulated}.}
\label{fig:mrclean}
\end{figure}

Finally, and for illustrative purposes, we show in Fig.~\ref{fig:simulated}
the simulated data set generated with the following: 13.13-day period,
FWHM=6.0 days and amplitude=0.1 with $\sigma_{\mathrm noise}=0.6$. The outputs
of {\sevensize PDM} and {\sevensize CLEAN}, are also shown. None of these
methods is able to detect the 13.13-day period. We show in
Fig.~\ref{fig:mrpdm} the outputs from {\sevensize MRPDM}. The simulated period
is detected, with its corresponding subharmonics, in the multiresolution
planes $w_2$ and $w_3$. The {\sevensize MRCLEAN} outputs, shown in
Fig.~\ref{fig:mrclean}, reveal a marginal detection of the 13.13-day period in
multiresolution plane $w_3$. We note that we would not consider this detection
as a true one when taken alone. However, since the same period is clearly
detected in two multiresolution planes of {\sevensize MRPDM}, we can establish
the existence of this period in the analysed data set.

\section{Conclusions}
\label{conclusions}

In this paper we have presented a multiresolution-based method for period
determination able to deal with unevenly sampled data. This constitutes a
significant improvement with respect to the wavelet-based method presented in
Paper~I, which is unable to deal with unevenly sampled data. The overall
performance of the present method is similar to the wavelet-based one, in the
sense that it allows us to detect superimposed periodic signals with lower
signal-to-noise ratios than in classical methods. We stress that one advantage
of the present method over classical methods is the simultaneous detection of
a period in more than one multiresolution plane, allowing to improve the
confidence of a given detection. Moreover, since here we are not forced to
lose or modify the information when averaging or interpolating the original
data, we can reach higher noise-to-signal ratios than in the wavelet-based
method described in Paper~I.

We note that the multiresolution decomposition scheme that we have used can be
interpreted as a particular case of scale-space filtering. In order to improve
isolation of periodic features, more general approaches could be used to
perform this decomposition. In this context, anisotropic diffusion schemes
proposed by \cite{perona} could be useful if properly tuned.

\section*{Acknowledgments}
We thank useful comments and suggestions from an anonymous referee.
We acknowledge partial support by DGI of the Ministerio de Ciencia y Tecnolog\'{\i}a (Spain) under grant AYA2001-3092, as well as partial support by the European Regional Development Fund (ERDF/FEDER).
This research has made use of facilities of CESCA and CEPBA, coordinated by C4 (Centre de Computaci\'o i Comunicacions de Catalunya).
XO is a researcher of the programme {\em Ram\'on y Cajal} funded by the
Spanish Ministery of Science and Technology and Centre de Visi\'o per Computador.
MR acknowledges support by a Marie Curie Fellowship of the European Community programme Improving Human Potential under contract number HPMF-CT-2002-02053.
MP is a researcher of the programme {\em Ram\'on y Cajal} funded by the
Spanish Ministery of Science and Technology and Universitat de Girona.

\bsp

\label{lastpage}


\begin{thebibliography}{1}


\bibitem[\protect\citeauthoryear{Baubaud et al.}{1986}]{baubaud}
Baubaud J., Witkin A., Baudin M., Duda R.O., 1986, IEEE Trans. Pattern Anal. Mach. Intell. PAMI-8, 26

\bibitem[\protect\citeauthoryear{Chui}{1992}]{chui}
Chui C.K., 1992, An Introduction to Wavelets, Boston Ac. Press, Boston 



\bibitem[\protect\citeauthoryear{Daubechies}{1992}]{daubechies}
Daubechies I., 1992, Ten Lectures on Wavelets, SIAM Press, Philadelphia

\bibitem[\protect\citeauthoryear{Foster}{1996}]{foster}
Foster G., 1996, AJ, 112, 1709

\bibitem[\protect\citeauthoryear{Jurkevich}{1971}]{jurkevich}
Jurkevich I., 1971, Ap\&SS, 13, 154 

\bibitem[\protect\citeauthoryear{Kaiser}{1994}]{kaiser} 
Kaiser G., 1994, A Friendly Guide to Wavelets, Birkh\"auser, Berlin

\bibitem[\protect\citeauthoryear{Lafler \& Kinman}{1965}]{lafler}
Lafler J., Kinman T.D., 1965, ApJS, 11, 216 


\bibitem[\protect\citeauthoryear{Levine et al.}{1996}]{levine}
Levine A.M., Bradt H., Cui W., Jernigan J.G., Morgan E.H., Remillard R., Shirey R.E., Smith D.A., 1996, ApJ, 469, L33

\bibitem[\protect\citeauthoryear{Lindeberg}{1990}]{lindeberg}
Lindeberg T., 1990, IEEE Trans. Pattern Anal. and Mach. Intell., 12, 234

\bibitem[\protect\citeauthoryear{Lomb}{1976}]{lomb}
Lomb N.R., 1976, Ap\&SS, 39, 447

\bibitem[\protect\citeauthoryear{Mallat}{1999}]{mallat}
Mallat S., 1999, A Wavelet Tour of Signal Processing, 2nd ed., Academic Press, San Diego

\bibitem[\protect\citeauthoryear{Meyer}{1993}]{meyer}
Meyer Y., 1993, Wavelets. Algorithm and Applications, SIAM Press, Philadelphia

\bibitem[\protect\citeauthoryear{Otazu et al.}{2002}]{otazu}
Otazu X., Rib\'o M., Peracaula M., Paredes J.M., N\'u\~nez J., 2002, MNRAS,
333, 365 (Paper~I)

\bibitem[\protect\citeauthoryear{Perona \& Malik}{1990}]{perona}
Perona P., Malik J., 1990, IEEE Trans. Pattern Anal. and Mach. Intell., 12, 629

\bibitem[\protect\citeauthoryear{Polygiannakis, Preka-Papadema \& Moussas}{2003}]{poly}
Polygiannakis J., Preka-Papadema P., Moussas X., 2003, MNRAS, 343, 725

\bibitem[\protect\citeauthoryear{Press \& Rybicky}{1989}]{press}
Press W.H., Rybicky G.B., 1989, ApJ, 338, 277

\bibitem[\protect\citeauthoryear{Rib\'o et al.}{2001}]{ribo}
Rib\'o M., Peracaula M., Paredes J.M., N\'u\~nez J., Otazu X., 2001, in Gim\'enez A., Reglero V., Winkler C., eds, Proc. Fourth INTEGRAL Workshop, Exploring the Gamma-Ray Universe. ESA Publications Division, Noordwijk, p. 333

\bibitem[\protect\citeauthoryear{Roberts, Leh\'ar \& Dreher}{Roberts et al.}{1987}]{roberts}
Roberts D.H., Leh\'ar J., Dreher J.W., 1987, AJ, 93, 968

\bibitem[\protect\citeauthoryear{Scargle}{1982}]{scargleA}
Scargle J.D., 1982, ApJ, 263, 835

\bibitem[\protect\citeauthoryear{Scargle}{1989}]{scargleB}
Scargle J.D., 1989, ApJ, 343, 874

\bibitem[\protect\citeauthoryear{Sporring et al.}{1997}]{sporring}
Sporring J., Niselsen M., Florack L., Johansen P., 1997, Gaussian Scale-Space Theory, Kluwer Ac. Pub., Drodrecht



\bibitem[\protect\citeauthoryear{Starck \& Murtagh}{2002}]{starck}
Starck J.L., Murtagh F., 2002, Astronomical Image and Data Analysis, Springer, Berlin

\bibitem[\protect\citeauthoryear{Stellingwerf}{1978}]{stellingwerf}
Stellingwerf R.F., 1978, ApJ, 224, 953
	
\bibitem[\protect\citeauthoryear{Szatm\'ary, Vink\'o \& G\'al}{Szatm\'ary et al.}{1994}]{szatmary}
Szatm\'ary K., Vink\'o J., G\'al J., 1994, A\&AS, 108, 377

\bibitem[\protect\citeauthoryear{Vetterli \& Kovacevic}{1995}]{vetterli}
Vetterli M., Kovacevic J., 1995, Wavelets and Subband Coding, Prentice Hall,
Englewood Cliffs, New Jersey

\bibitem[\protect\citeauthoryear{Witkin}{1994}]{witkin94}
Witkin A.P., 1984, Image Understanding 1984, Ullman and Richards editors, Norwood Ablex, New Jersey




\end{thebibliography}
\end{document}